\magnification=1500 \tolerance=1000

\centerline{\bf Neutrinos and Their Charged Cousins:}
\centerline{\bf Are They Secret Sharers?}
\centerline{\it Sheldon Lee Glashow}
\centerline{\it Department of Physics}
\centerline{\it Boston University}
\centerline{\it Boston, 02215 MA}

{\medskip\narrower\noindent \sl Masses and mixings of quarks  and leptons
differ wildly from one another.
Thus it is all the more challenging to search for some  hidden attribute 
that they may share\medskip}

``Neutrinos they are very small. They have no charge and have no mass
and do not interact at all.'' So wrote 
the renowned author John Updike, erring twice
in one couplet.  Indeed, neutrinos are the most storied of
particles. Think of Pauli's famous letter proposing them as a
``desperate remedy'' for the apparent failure of energy conservation
in beta decay. He waited  25 years for his neutrinos to be observed.
 Think of Ray Davis' good news and bad: that he had
succeeded in detecting solar neutrinos, but that they were too
few. And only once did I witness a standing ovation at a physics
conference --- at Neutrino-98 in Takayama, Japan, where scientists at
Super-Kamiokande announced the observation  of atmospheric neutrino
oscillations, an effect I had anticipated almost two decades
earlier.\footnote{$^1$}{S.L. Glashow {\it in\/} Quarks \& Leptons,
Carg\`ese 1979 (Plenum, 1980, New York) pp. 687-713, see p. 707.}  A
few years later scientists at SNO, the Sudbury Neutrino Observatory
in Canada,
announced their triumphant resolution of Ray's decades-old solar
neutrino problem. More   front-page neutrino
stories are  sure to follow!

\medskip Neutrinos, once  ghostlike and elusive, are now
mundane. They are seen and studied  from many sources:
reactors,  accelerators, radioactive decays, Earth's
interior, cosmic-ray interactions in the stratosphere and from two
stars: the sun and supernova 1987a.  What we've learned about them is
succinctly described by three left-handed neutrino states and a
relatively small number of adjustable parameters, but many vexing
questions remain: \medskip

Is the neutrino mass spectrum normal or inverted?
Are the masses  lepton-number violating, {\it i.e.,} Majorana; or
 lepton-number conserving, {\it i.e.,} Dirac;
  or are they a bit of both,
thereby putting
 light singlet states, so-called sterile neutrinos,  into play?
 Will astrophysical observations or the
Katrin experiment or searches for neutrinoless double beta decay,
determine the neutrino mass scale?  Is there observable
CP violation in the neutrino sector?  Will future data belie our
simple model? Let me put aside these questions  and begin my tale
 with a brief historical digression. Recall the
 five eras of lepton hadron symmetry: 
\medskip

\item{$(1\times 1)^2$} During the 1920s atoms and their nuclei seemed
to be made up from  just two elementary constituents: protons and
electrons --- one hadron and one lepton, although neither word had yet
been coined.  \smallskip

\item{$(1\times 2)^2$} 
The discovery  of neutrons and the invention of neutrinos
led to the second era: one doublet of nucleons and one of leptons.
\smallskip

\item{$(1\times 3)^2$} The detection  of strange particles
and muons in cosmic rays led Bob
Marshak to propose the Kiev symmetry in 1959. The fundamental Sakata
triplet of hadrons ($p,\,n,\,\Lambda$) was likened to the lepton
triplet  ($\nu, e,\,\mu$). This era was short lived.
Sakata's triplet was replaced by  three  quarks, but 
the discovery of a second neutrino in 1963 undid 
Marshak's symmetry.\smallskip

\item{$(2\times 2)^2$} Soon after Gell-Mann and Zweig devised quarks,
James Bjorken and I proposed  the  existence of yet one more. Our
reasoning was purely aesthetic: with charm, two quark doublets 
would accompany  the two  lepton doublets, thus restoring
lepton-hadron symmetry.
Six years passed before John Iliopoulos, Luciano Maiani and I offered
substantive and convincing arguments for the existence of
charm, another four before the experimental discovery of charmonium.
\vfill\eject

\item{$(3\times 2)^2$} Mere months after the discovery of
the $J/\Psi$, groups led by Marty Perl and Leon Lederman spotted half
the members of a third family of fundamental fermions. Top quarks
and tau neutrinos would show up  decades later. In the current
and longest-lived era of lepton-hadron  symmetry, there are three
doublets  of quarks and and three of leptons.\medskip

Fermion masses and mixings were much simpler in the two-doublet
era than now. Back then, with only two families and no evidence
for neutrino masses, the `flavor problem'  involved only seven
parameters: four quark masses, two charged lepton masses and 
Cabibbo's angle. But when $2\times 2$ matrices became $3\times 3$,
things got  complicated.
  Quark masses and
mixings involve six masses and four 
Cabibbo-Kobayashi-Maskawa  (CKM) parameters.
Meanwhile, observations of solar and atmospheric 
oscillations showed that  neutrinos   have small but consequential
masses. Thus the lepton sector involves  ten  analogous quantities:
six lepton masses and four Pontecorvo-Maki-Nakagawa-Sakata (PMNS)  
parameters, as well as  two  infuriatingly
inaccessible  Majorana phases.\footnote*{Here  we
assume there to be three relevant neutrino states with
lepton number violating (Majorana) masses.}
 All twenty of the flavor parameters are either measured or  
constrained. And yet, frustratingly, no significant relationship among
them strikes the eye nor has any been deduced from a plausible
theoretical framework.  \medskip

The CKM matrix has little in common with its leptonic analog.
All three quark mixing angles are small: Cabibbo's is about
$13^\circ$, the others much smaller.  Contrariwise, atmospheric
neutrino oscillations seem nearly maximal and the solar oscillation
angle is large as well. But could there be some common attribute
hiding amongst the masses of quarks and leptons, if not amongst their
mixings? 
\medskip

The three charged leptons are widely disparate in mass.
So  are  the three up-type quarks and the three down-type
quarks. Denote  the masses
 in each
category by $B$ (for biggest), $M$ and $S$ (for smallest).
  We consider several shared measures  of their disparity
which might also characterize  neutrino masses. In order of descending
strength, they are:

\medskip

\item{ F1:} For  each category (charged leptons, up-type quarks and
down-type quarks)  the  ratios $B/M$ and $M/S$ are both 
large, ranging from 17 to 140. Here we suppose the
corresponding neutrino mass ratios to be greater than ten.

\medskip

\item{ F2:} Here we impose a weaker constraint, requiring  only one 
neutrino mass ratio, $B/S$,  to exceed ten.
\medskip

\item{ F3:} An even weaker indicator of mass disparity is  for the
three masses  not to form a triangle or equivalently, for $B$
to exceed $M+S$.
Charged lepton masses easily satisfy this inequality, as do
those of quarks of either charge. Here we assume the same 
for neutrino masses.
\medskip

\noindent F1 implies F2.   
For  the known values of $\Delta_s$ and $\Delta_a$,  
F2 implies F3.
We exhibit the implications of each feature
on  three observables: $\Sigma$, the sum of the absolute values of
the neutrino  masses (which may be determined astrophysically);
$m_\beta$, the effective mass of the electron neutrino
(which may be determined from studies of the beta-decay endpoint);
and $m_{\beta\beta}$, the $e$-$e$ element of the neutrino mass matrix
(which may be determined from the rate of neutrinoless double beta decay).
In a conventional notation, the three neutrino  observabkles are given by:
$$\eqalign{  
\Sigma & \equiv |m_1|+|m_2|+|m_3|\,,\cr
m_\beta & \approx \big(c_3^2|m_1|^2+s_3^2|m_2|^2\big)^{1/2}\,,\cr
m_{\beta\beta} & \approx |c_3^2m_1+s_3^2m_2|\,,\cr}$$
where $\Sigma>m_\beta\ge m_{\beta\beta}$. We  use the following 
experimental data
for our analyses: $s_3^2\equiv\sin^2{\theta_{12}}\approx
0.3$, $s_2^2\equiv \sin^2{\theta_{13}}\le 0.03$,
$|m_3^2-m_1^2|\approx 2400$~meV$^2$ and $m_2^2-m_1^2\approx 77$~meV$^2$.

\medskip

Should neutrino masses share feature F1 with their charged
counterparts, one neutrino must be nearly massless and the hierarchy
must be normal. We obtain (with masses in meV):
$$\eqalign{
 {\rm F1\ (normal):} \quad &\Sigma\approx 58,\ \ m_\beta\approx 5, \ \ 
m_{\beta\beta}\approx 3,\cr
{\rm F1\ (inverted):} \quad & {\rm Not\ possible.}\cr}$$

Should they share feature F2, one neutrino must be nearly
massless, but the hierarchy may be either normal or inverted. For the
normal case, the consequences are the same as for  F1. 
We obtain:
$$\eqalign{
 {\rm F2\ (normal):} \quad &\Sigma\approx 58,\ \ m_\beta\approx 5, \ \ 
m_{\beta\beta}\approx 3,\cr
 {\rm F2\ (inverted):}\quad &\Sigma\approx 98,\ \ m_{\beta\beta}\le m_{\beta}
\approx 49.}$$

Should they share feature F3, the hierarchy may be either
normal or inverted and we obtain:
$$\eqalign{
 {\rm F3\ (normal):} \quad &\Sigma\approx 113,\ \ 
m_{\beta\beta}< m_\beta \approx 28, \cr 
 {\rm F3\ (inverted):}\quad &\Sigma\approx 98,\ \ m_{\beta\beta}\le m_{\beta}
\approx 49.}$$
Even this weakest indication of  neutrino mass disparity 
severely constrains the three neutrino
observables. If it is satisfied,  neither  
endpoint effects in tritium decay nor neutrinoless double beta decay
are likely to be observed   in the forseeable future.

\medskip Conversely, either a measurement of the electron neutrino mass, or
a detection of neutrinoless double beta decay,
or a convincing astrophysical argument that 
 $\Sigma$  much exceeds 100 meV 
 would imply
 that none  of the above  features of  charged
 fermion masses  characterize neutrino masses.
 They
would have  to differ from those of quarks and charged leptons in
virtually
 every imaginable fashion: in their magnitudes and mixing parameters, 
 their degree of disparity, as well as  the manner of their origin. 
\medskip

Suppose the Katrin experiment finds the electron neutrino mass
to exceed 200~meV (the lower limit of its 
sensitivity)\footnote{$^2$}{See: {\tt www-ik.fzk.de/tritium/}.}.
Then  the three neutrino masses would not
only form a triangle, but one which is equilateral to a precision of
3\% or better!  Thus are we led a fourth and quite different feature
which could  be shared by  quark and lepton mass matrices:
\vfill\eject

\item{F4:} All three quark mixing angles are small and
 the off-diagonal entries of $M_d$ are small in the up-quark
basis.  Here we assume the latter property to characterize $M_\nu$ as well.
 Although two  neutrino mixing angles are {\bf not}
small, the off-diagonal entries of
$M_\nu$ will be  small in the charged lepton basis if and only
if the three
neutrino masses  are nearly degenerate.  \medskip

\smallskip\noindent  Experimenters should be pleased
were neutrinos to share this feature.
All three neutrino observables would
 be relatively large, satisfying: 
$$m_{\beta\beta}\approx m_\beta\approx \Sigma/3 \ge 100\ \rm meV.$$
The  inequality enforces neutrino mass degeneracy to within 10\%, thus
  ensuring   the
 diagonal entries of $M_\nu$ to be  nearly equal and its 
off-diagonal entries to be  at least an order of magnitude 
smaller.  Thus F4 requires $M_\nu$ to have
a simple ``One plus Zee\footnote{$^3$}{A. Zee, Phys. Lett. 93B (1980) 389.}''
form --- 
 a multiple of the unit
matrix augmented by three far smaller (but non-zero) off-diagonal 
entries.\footnote{$^4$}{S.L. Glashow, arXiv: 0912.4976 [hep-ph].}
But can  anyone suggest a sensible  theoretical scheme  yielding  this 
most pleasing of textures?

\bye

\bye